\documentstyle[epsf,twocolumn,prl,aps]{revtex}
 \newcommand{\figwidth}{3.0in} 

\newcommand{\tj}{$t$-$J$}

\newcommand{\etal}{{\it et al}}
\newcommand{\ij}{\langle ij\rangle}
\renewcommand{\S}{{\vec S}}  

\begin{document}
\draft

 \twocolumn[\hsize\textwidth\columnwidth\hsize\csname @twocolumnfalse\endcsname

\title{DMRG Study of the Striped Phase in the 2D $t$-$J$ model}
\author{ Steven R.\ White$^1$ and D.J.\ Scalapino$^2$}
\address{ 
$^1$Department of Physics and Astronomy,
University of California,
Irvine, CA 92697
}
\address{ 
$^2$Department of Physics,
University of California,
Santa Barbara, CA 93106
}
\date{\today}
\maketitle
\begin{abstract}
\noindent 

Using the density matrix renormalization group (DMRG),
we study the 2D \tj\ model at a hole doping of $x={1\over8}$ on
clusters as large as $19 \times 8$.
We find a striped phase which is consistent with recent neutron
scattering experiments. We find that bond-centered and
site-centered stripes have nearly the same energy,
suggesting that in the absence of pinning effects the domain
walls can fluctuate. 

\end{abstract}

 ]

In the low temperature tetragonal (LTT) phase of 
La$_{1.6-x}$Nd$_{0.4}$Sr$_x$CuO$_4$, the tilt pattern of the
CuO$_6$ octahedra form lines of displaced oxygens parallel to the
Cu-O bond directions. These lines are rotated by 90$^\circ$ between
adjacent layers. At a filling of $x={1\over8}$, superconductivity is
suppressed and neutron scattering studies\cite{TN,T} reveal a striped domain
wall ordering of holes and spins which is believed to be
commensurately locked by the tilt distortion of the lattice. One
model for this striped order\cite{TN,T} is illustrated in Fig.~1(a).
Here the charge domain walls are shown running vertically and
centered along the Cu-O-Cu legs, although the phase information required
to determine whether the domains should be leg centered or bond centered
(centered between two legs) is not known. As shown, the domains are
separated by four Cu-O-Cu spacings and for $x={1\over8}$ contain
one hole per two $4\times 1$ domain wall unit cells. 
This latter feature is at odds with one-electron Hartree-Fock
calculations\cite{hfdomain} which predict a domain wall filling of one
hole per domain wall unit cell. The spins in the regions between the
walls are antiferromagnetically correlated with a $\pi$ phase shift
across a domain wall.
When $x\neq{1\over8}$, superconductivity is found to
coexist with a weakened domain wall ordering, suggesting a close
connection between the two.

Here we present results of numerical density matrix renormalization
group (DMRG)\cite{dmrg} calculations 
for a \tj\ model with a hole doping $x={1\over8}$.
We find evidence for domain walls with $\pi$ phase-shifted
antiferromagnetic regions separating the walls, and with a filling
of one hole per two $4\times 1$ domain wall unit cells.
Kivelson and Emery\cite{K} have suggested that domain walls
arise when phase separation of the holes into uniform hole-rich
and hole-poor regions is frustrated by 
long-range Coulomb forces.  
The question of whether, in fact, the \tj\ model exhibits phase 
separation for the relevant physical values of
$J/t$ and doping $x$ remains controversial \cite{nophasesep,phasesep}.
Our present results show that
long-range Coulomb forces are not necessary for the formation
of domain walls. 

Depending on the dimensions and boundary conditions (BCs) of the 
cluster we study, the domain walls 
may be 
site-centered, 
as shown in Fig. 1(a), bond centered, or in 
\begin{figure}[ht]
\epsfxsize=3.375 in\centerline{\epsffile{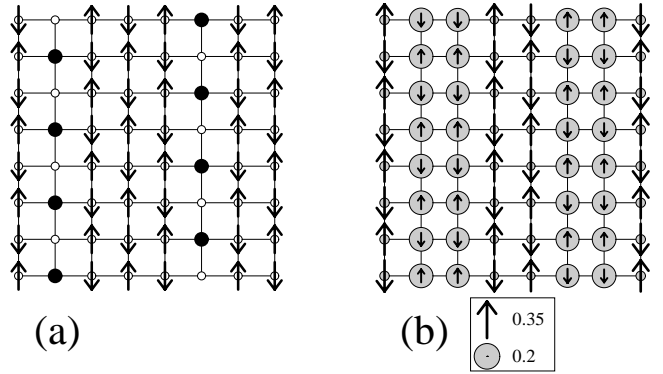}}
\caption{(a) Spin and hole structure suggested in 
Ref. [1] to
account for neutron scattering experiments.
(b) Hole density and spin moments for the central $8\times8$
region of a $16\times8$ \tj\ system.
The diameter of the gray holes is proportional to the hole density
$1-\langle n_i\rangle$,
and the length of the arrows is proportional to $\langle S^z_i \rangle$,
according to the scales shown.
}
\end{figure}
\noindent between. 
In contrast to Fig. 1(a), however, the site
centered domain walls have substantial hole densities over three
rows of sites, rather than one.
Previous attempts to understand the charge degrees of freedom of
the striped phase have focused
on one dimensional models\cite{NW,K}. We consider another approach, in
which coupled ladders are used to model the 2D system.
In particular, in order to understand bond-centered stripes,
we consider an array of two-leg ladders which are
coupled antiferromagnetically via a mean field. We find that the
$\pi$-phase-shifted magnetic order of the bond-centered
striped phase can be understood 
within this mean field picture.
Similarly, the magnetic order of the site-centered striped phase
can be understood in terms of doped three-leg ladders coupled
antiferromagnetically to undoped chains.

The \tj\ Hamiltonian in the subspace of no doubly occupied sites
is given by
\begin{equation}
H = - t \sum_{\langle ij \rangle s}
      ( c_{is}^{\dagger}c_{js}
                + {\rm h.c.}) + 
J \sum_{\langle ij \rangle}
      ( {\bf S}_{i} \! \cdot \! {\bf S}_{j} -
         \frac{n_i n_j}{4} ) .
\label{tj-ham}
\end{equation}
Here $\ij$ are near-neighbor sites, $s$ is a spin index, $\S_i =
c^\dagger_{i,s} {s}_{s,s'}c_{i,s'}$ and $n_i=
c^\dagger_{i\uparrow}c_{i\uparrow} +
c^\dagger_{i\downarrow}c_{i\downarrow}$, with $c^\dagger_{is}$
($c_{is}$) an operator which creates (destroys) an electron at site $i$
with spin $s$.
The near-neighbor hopping interaction is $t$ and the near-neighbor
exchange interaction is $J$.
We refer to the Cu-Cu lattice spacing as $a$
and measure energies in units of $t$. We consider only $J/t=0.35$ here.

We present results here for $L\times8$ clusters, with $L$ as
large as 19. As first discussed by Liang and
Pang\cite{liangpang}, the truncation errors in a DMRG calculation
typically rise {\it exponentially}  with the width of a
large two-dimensional system (while only linearly with the
length). However, the errors also tend to {\it fall}  exponentially with
the number of states kept per block. Consequently, while studies of doped
$L\times8$ clusters are quite difficult, by keeping from
1000-2000 states per block, we can obtain useful results, with
truncation errors of $0.0002$ - $0.0001$.
We are able to keep this many states because of recent
improvements in the DMRG finite-system algorithm\cite{cavo}.

The nature of the ground state of the 2D $t$-$J$ systems causes
additional numerical difficulties. Rather than an approximately
homogeneous phase, we find that the system tends
to have inhomogeneous charge and spin distributions (such as domain
walls), which can be pinned by the open BCs
usually used in DMRG. Usually more than one such low-energy configuration
is possible: for example, one could have horizontal as opposed to
vertical stripes. A DMRG calculation involves sweeps through the sites
of the lattice, and the energy of the approximate DMRG ground state 
of the system is decreased mostly through ``local'' improvements of the
wavefunction.  We find that in a large 2D system DMRG is usually unable
to tunnel between two substantially different low-energy configurations. 
Even when a low energy tunneling path exists between two very different
configurations, the calculation may move along the path slowly.
To deal with these difficulties, we usually perform several
simulations for each system. These systems differ in the
charge and spin configurations in the first few sweeps. Later
sweeps drive the system to a local energy minimum. One can
then compare the total energy of different simulations to find
which configuration is the ground state.  The charge
and spin configurations can be controlled in two ways: 1) by
adjusting the total quantum numbers of the system at each step
as the lattice is first built up from a few sites; and 2) by
applying local chemical potentials and magnetic fields for the
first few sweeps. Unfortunately, it is possible to miss the true
ground state configuration if it is substantially different from
what one expects. However, unlike an ordinary variational calculation,
only the crudest overall features of the wavefunction, such as
the general location of the domain walls, are specified in the 
initial sweeps.
These various runs can give
substantial insight into what sorts of low-energy configurations 
can possibly occur under slightly different BCs
or small perturbations to the Hamiltonian.

Figure 1(b) shows the charge and spin density in the ground state for
the central $8\times8$ section of a
$16\times8$ system with $J/t=0.35$ and 16 holes, corresponding to a
filling $x={1\over8}$. Periodic BCs were used
in the $y$ direction, and open BCs in the
$x$ direction. Along the left and right edges of the system a 
small staggered magnetic field of $0.1t$ was applied. The 
BCs and the edge staggered field serve to orient and pin the domain
walls in the configuration shown. In an LTT phase, the domain walls
are oriented, and possibly pinned, by the lattice distortion. 
The staggered edge field further
acts to pick a direction for the spin order, which allows direct
measurement of the spin configurations and reduces truncation errors
in the DMRG calculation. Previous to this calculation, dozens of
simulations were performed, mostly on $8\times8$ clusters, to
find the nature of the ground state and the effect of various
BCs.  Included were several initial conditions 
corresponding to phase separation, with the hole cluster 
either on the edge or in the center of the system. 
These phase-separated configurations were unstable, with the
hole cluster tending to split or lengthen into domain walls.
A single eight-hole vertical domain wall was also unstable,
even when initial conditions and boundary staggered magnetic
fields favored one. Objects resembling diagonal domain walls
have been observed in three chain and four chain
ladders\cite{threechain,fourchain}, but attempts to stabilize
a diagonal domain wall on an $8\times8$ system instead yielded
a bent domain wall with the central part aligned in the (1,0)
direction.
Periodic BCs in the $y$ direction tend to
favor vertical domain walls; open BCs in the
direction of the stripes tend to suppress them.

In the simulation shown in Fig 1(b), eleven sweeps were 
performed, and in the final sweep 1400 states were kept. A local chemical
potential was applied to confine the holes to the width-two
stripes shown for the first six sweeps, and then removed. No initial 
magnetic field was needed away from the left and right edges 
to orient the $\pi$-shifted antiferromagnetic domains as shown.

Figure 2 shows the domain wall structure in a different 
\begin{figure}[hb]
\epsfxsize=\figwidth\centerline{\epsffile{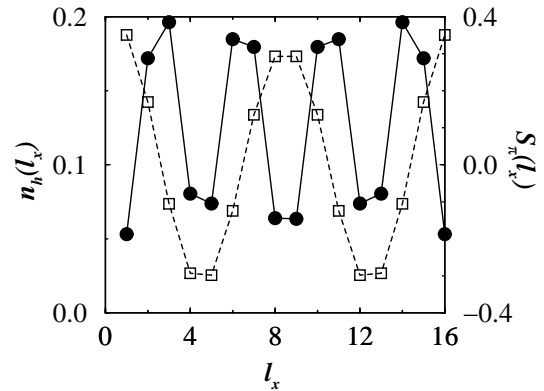}}
\caption{
Average hole density $n_h(\ell_x)$ (solid circles) and
spin structure function $S_\pi({\ell_x})$  (open squares)
for the $16\times8$ system of Fig. 1(b).
}
\end{figure}
\noindent way.
With the solid circles we show the local hole density 
$n_h(\ell) = 1 - <c^\dagger_{\ell\uparrow}c_{\ell\uparrow} +
c^\dagger_{\ell\downarrow}c_{\ell\downarrow}>$
as a function of the $x$-coordinate $\ell_x$. 
The bond-centered nature of these domain walls is clearly evident. 
To show the spin structure, we define
\begin{equation}
S_\pi(l_x) = 1/L_y \sum_{l_y = 1}^{L_y} (-1)^{l_x+l_y} 
\langle S_z(l_x,l_y) \rangle.
\end{equation}
With the open squares, we show $S_\pi(l_x)$. 
The period-8 spin structure is clearly evident.

The boundary conditions have a strong effect on the structure of 
the domain walls which appear. Bond-centered domain walls tend
to form one lattice spacing away from an open boundary.
This initially led us to believe that site-centered domain walls
were not stable, but subsequent simulations showed that site
centered walls occur also.  In Fig. 3 we show the local hole
density and spin structure function $S_\pi(l_x)$ for a
$19\times8$ cluster with 20 holes. The same BCs and edge
magnetic field as for the system shown in Fig. 2 were applied.
A system such as this with an odd number of domain walls 
and open BCs in the $x$ direction
is forced by symmetry under reflection about a vertical line to
have a site-centered domain wall in the center if $L_x$ is odd,
and a bond-centered wall if $L_x$ is even. In the calculation
shown, reflection symmetry is used explicitly, which ensures that
a site centered domain wall appears in the center. Note that 
the second and fourth domain walls, which are not so constrained by
geometrical effects, are more site-centered than bond centered.
We have compared the local energies averaged over $4\times4$ regions covering
site-centered and bond-centered walls; the difference in energy
per site between these was within our numerical errors for
local energies, with both giving $E/N \approx 0.62t \pm 0.01t$. 
In addition to bond centered and site
centered walls, asymmetrical walls can occur. The close energy
differences between these different types of walls suggests
that a large 2D \tj\  system at $x=1/8$ might have fluctuating 
domain walls.
\begin{figure}
\epsfxsize=\figwidth\centerline{\epsffile{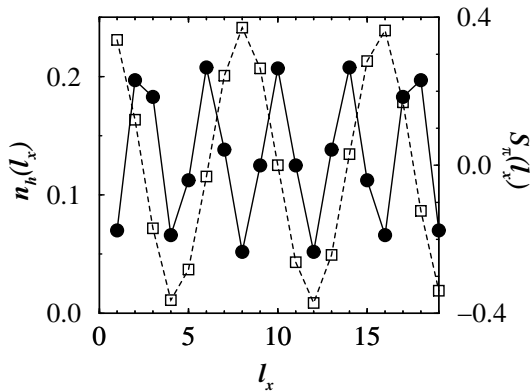}}
\caption{
Average hole density $n_h(\ell_x)$ (solid circles) and
spin structure function $S_\pi({\ell_x})$ (open squares)
for a $19\times8$ system.
}
\end{figure}

The $\pi$-phase-shifted antiferromagnetic
regions reduce the energy for transverse hopping of  holes within
a domain wall.
To understand in more detail the bond-centered striped structure, 
we consider a model of antiferromagnetically coupled two-leg ladders. 
Ladders doped with $x=0.25$ are alternated with undoped ladders,
and no hopping is allowed between ladders. 
Ladders are exchange-coupled
via a mean field, which is staggered along a ladder, but which may or may
not have a $\pi$ phase shift across a doped ladder. The properties
of a single ladder are calculated with DMRG, with a static magnetic field
with wavevector ($\pi$,$\pi$) or ($0$,$\pi$). 
In Fig. 4 we show the
magnetic response $|\langle S_z \rangle|$ to an applied field with
magnitude $h$. As expected, an undoped ladder has a much greater
response at the N\'eel wavevector ($\pi$,$\pi$). A doped ladder,
in contrast, shows a substantially greater response at ($0$,$\pi$).
Hence the mean field treatment shows the $\pi$ phase shift seen
in the 2D calculations. The mean field self-consistency conditions
are
\begin{equation}
h_{d,u} = J |\langle S_z \rangle_{u,d}|
\end{equation}
where $u$ and $d$ stand for doped and undoped ladders.
From the results shown in Fig. 3, we find 
$|\langle S_z \rangle_u| = 0.32$ and 
$|\langle S_z \rangle_d| = 0.15$. The results from the $16\times8$
system, in contrast, are 
$|\langle S_z \rangle_u| = 0.29$ and 
$|\langle S_z \rangle_d| = 0.13$. As one expects, the mean field
treatment overestimates the magnetic order. (We expect that correction for 
truncation errors and finite size effects would further decrease 
the DMRG results.) The energy of this mean-field striped 
\begin{figure}
\epsfxsize=2.5 in\centerline{\epsffile{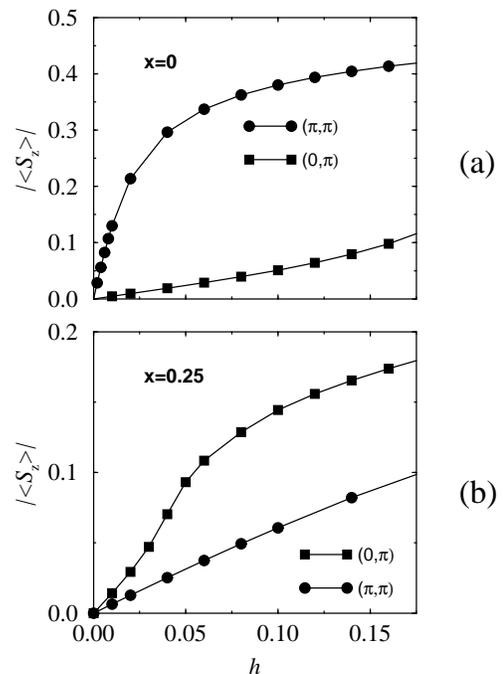}}
\caption{
Magnetization per site $|\langle S_z \rangle|$ 
induced by an applied magnetic field $h$ at wavevectors
$(\pi,\pi)$ and $(0,\pi)$ on a $2\times32$ ladder.
(a) An undoped ladder. (b) A ladder with doping $x=0.25$.
}
\end{figure}
\noindent 
phase (with $x=0$ and $x=0.25$), including the exchange
coupling between ladders, is about 2\% higher 
than the energy of an array of uncoupled ladders at uniform
density ($x=0.125$).
Hence the mean field approach does not predict the charge ordering of
the striped phase.
Nevertheless, these
results suggest that coupled ladders are natural starting points 
for understanding striped phases.

A similar mean field treatment can be made for site-centered
domain walls, coupling doped three-leg ladders with undoped
single chains. This also yields $\pi$-phase-shifted
antiferromagnetism, with reasonable magnitudes for 
$|\langle S_z \rangle|$.  We will present these results elsewhere.

According to the Maxwell construction, 
the occurence of phase separation results in a linear dependence
of the energy on the filling over a range of fillings. While there
has been disagreement in previous studies about whether 
phase separation occurs in the low-doping region for $J/t=0.3-0.5$, 
it is clear that
the curvature in the energy versus filling curve is 
small\cite{phasesep,nophasesep}. 
The possibility of a striped phase, which has generally not been
considered in these studies, makes the
analysis more difficult. In particular, if we assume that at
low doping, holes go into a single domain wall, then it would
appear from the energy alone
that one has phase separation up to a filling of $\sim N^{-1/2}$,
where $N$ is the number of sites in the system. At higher
dopings, an array of weakly repulsive, widely spaced 
domain walls would show a nearly linear energy versus
filling dependence. In fact, we have observed exactly this
scenario on a $12\times 6$ system at $J/t=0.5$, with periodic
BCs in the $y$ direction and open BCs in the $x$ direction.
Two holes bind into a pair, with a binding energy of 
$0.26t \pm 0.01t$. Two pairs bind into one vertical domain wall, with
binding energy $0.10t \pm 0.03t$. 
These domain walls are very similar to the ones seen in
the $L\times8$ systems, with either site-centered or
bond-centered walls possible, but with a linear hole density greater
by a factor of $4/3$.
Eight holes form two widely spaced domain walls; twelve holes
form three domain walls. We will report on these results in more
detail elsewhere.

Why is superconductivity suppressed specifically at $x=1/8$ in 
La$_{1.6-x}$Nd$_{0.4}$Sr$_x$CuO$_4$? More work is needed to
answer this question, but we can make some general statements.
First, viewing the stripes as coupled two-leg ladders,
there does not appear to be any anomalous feature in an
isolated two-leg ladder at $x=0.25$, such as a charge gap, which would
suggest that domain walls must occur with exactly this doping.
In any case, the charge on the domain walls spills out onto the rows
of adjacent sites; our bond-centered domain walls have a hole
density of $x \approx 0.18$ on the walls and $x \approx 0.07$ on
the adjacent sites.
This suggests, along with our results on the $12\times 6$ system,
that domain walls can occur with a range of dopings. 

The period of charge density wave (CDW) correlations, however, 
is insensitive to the broadening of the walls. We have
observed this effect on width-four models of domain-walls.
The CDW can be
viewed as a one dimensional line of hole pairs. The pairs extend 
beyond the two-site width of the wall, but the period is set by
the one-dimensional hole-density.
This period is 4$a$ at $x=0.25$, the same as the
transverse period of the stripes.  
In the LTT phase, the CuO$_6$ tilt structure causes the domain walls to
be perpendicular in adjacent planes\cite{TN}. This suggests that
a coupling between planes, such as through an 
electrostatic potential\cite{K} or through a lattice distortion, 
could induce a static CDW order along the domain
walls. This CDW order would tend to suppress superconductivity.


We would like to thank J.~Tranquada for discussions of his experimental
results, and J. Lawrence and S. Kivelson for helpful discussions.  
SRW acknowledges support from the NSF under 
Grant No. DMR-9509945, and DJS acknowledges support from the
NSF under grant numbers PHY-9407194 and DMR-9527304.

\newpage


\begin{references}
\bibitem{TN}J.M.~Tranquada \etal, Nature {\bf 375}, 561 (1995);
\prb {\bf 54}, 7489 (1996).

\bibitem{T}J.M.~Tranquada et. al., \prl {\bf 78}, 338 (1997).

\bibitem{hfdomain}J.~Zaanen and O.~Gunnarsson, \prb {\bf 40}, 7391 (1989);
D.~Poilblanc and T.M.~Rice, \prb {\bf 39}, 9749 (1989);
H.J.~Schulz, J.~Physique, {\bf 50}, 2833 (1989);
K. Machida, Physica C {\bf 158}, 192 (1989);
K. Kato, K. Machida, H. Nakanishi and Fujita, J. Phys. Soc. Jpn.
59, 1047 (1990);
J.A.~Verg\'es \etal, \prb {\bf 43}, 6099 (1991);
M.~Inui and P.B.~Littlewood, \prb {\bf 44}, 4415 (1991);
J.~Zaanen and A.M.~Oles, Ann.\ Physik {\bf 5}, 224, (1996).

\bibitem{dmrg} S.R. White, \prl {\bf 69}, 2863 (1992),
\prb {\bf 48}, 10345 (1993).

\bibitem{K}S.A.~Kivelson and V.J.~Emery, p.~619 in {\it Proc.
``Strongly Correlated Electronic Materials: The Los Alamos Symposium
1993,''} K.S.~Bedell \etal, eds. (Addison Wesley, Redwood City, Ca.,
1994); S.A.~Kivelson and V.J.~Emery, preprint (cond-mat/9603009).

\bibitem{nophasesep} 
M.U. Luchini, {\em et al.},
Physica C {\bf 185-189},  141  (1991);
W.O. Putikka, {\em et al.}, \prl {\bf 68},  538
(1992); H. Fehske, {\em et al.}, \prb {\bf 44}, 8473  (1991);
D. Poilblanc, \prb {\bf 52},  9201 (1995);
R. Valenti and C. Gros,  \prl {\bf 68}, 2402 (1992);
H. Yokoyama and M. Ogata, J.\ Phys.\ Soc.\ Japan {\bf 65}, 3615 (1996);
M. Kohno, \prb {\bf 55},  1435 (1997).


\bibitem{phasesep} C. S. Hellberg and E. Manousakis,
(cond-mat/9611195); V.J. Emery, {\em et al.},
\prl {\bf 64}, 475 (1990).

\bibitem{NW}C.~Nayak and F.~Wilczek, \prl {\bf 78}, 2465 (1997).

\bibitem{liangpang}  S. Liang and H. Pang, \prb {\bf 49}, 9214
(1994).

\bibitem{cavo} S.R. White, \prl {\bf 77}, 3633 (1996).

\bibitem{threechain}S.R.~White and D.J.~Scalapino, unpublished.

\bibitem{fourchain}S.R.~White and D.J.~Scalapino, to appear
in PRB  (cond-mat/9608138). 



\end{references}
\end{document}